\documentclass[11pt,british]{article}
\usepackage{libertine}

\usepackage{euler}
\usepackage{textcomp}
\usepackage{titlesec}
\usepackage{textcase} 
\usepackage{soul}
\usepackage{amsmath}
\sodef\allcapsspacing{\upshape}{0.25em}{0.8em}{0.6em} 

\titleformat*{\section}{\color{gray}\Large\sc}
\titleformat*{\subsection}{\color{gray}\large}
\titleformat*{\paragraph}{\bfseries}

\usepackage{tcolorbox}
\tcbset{
  coltitle=black,fonttitle=\bfseries,
  colback=blue!7!white,colframe=gray!25!white,width=1\textwidth,valign=center,
before=\par\bigskip\centering,after=\par}

\usepackage{colordvi}
\usepackage{epsf}
\usepackage{graphicx}
\usepackage{tikz}
\usepackage{xcolor}
\usepackage{hyperref}
\usepackage{setspace}
\usepackage[numbers,sort&compress]{natbib}
\usepackage[english=american]{csquotes}

\usepackage[total={6.6in, 9.2in}]{geometry}
\definecolor{cadmiumgreen}{rgb}{0.0, 0.42, 0.24}

\newcommand{\gray}[1]{\textcolor{gray}{#1}}

\usepackage{amssymb}

\newcommand{\e}{\mathrm{e}}
\newcommand{\norm}[1]{\|#1\|}
\newcommand{\one}{\mathbf{1}}

\newcommand{\deff}{d^{\mathrm{eff}}}
\newcommand{\av}[1]{\langle #1 \rangle} 
\newcommand{\tr}{\textrm{tr}}

\begin{document}


\title{\LARGE \sc{\gray{Equilibration times in closed quantum many-body systems}}}

\author{\normalsize{H. Wilming$^{1,2}$, T. R. de Oliveira$^3$, A. J. Short$^{4}$, J. Eisert$^{1}$}}

\date{}
\maketitle
\vspace*{-.8cm}
\begin{center}
{\footnotesize 1 Dahlem Center for Complex Quantum Systems, Freie Universit{\"a}t Berlin, 14195 Berlin, Germany\\
2 Institute  for  Theoretical  Physics,  ETH  Zurich,  8093  Zurich,  Switzerland \\
3 Instituto de F\'isica, Universidade Federal Fluminense, 24
210-346, Niter\'oi, RJ, Brazil\\
4 H. H. Wills Physics Laboratory, University of Bristol,
Tyndall Avenue, Bristol, BS8 1TL, UK\\}
\end{center}


\begin{abstract}
For a quantum system to be captured by a stationary statistical ensemble, as is common in
thermodynamics and statistical mechanics, it is necessary that it reaches some apparently stationary state in the first place.
In this book chapter, we discuss the problem of equilibration and specifically provide insights into 
how long it takes to reach equilibrium in closed quantum systems. We first briefly discuss the connection of this problem with recent experiments and forthcoming quantum simulators. Then we provide a comprehensive discussion of equilibration from a heuristic point of view, with a focus on providing an intuitive understanding and connecting the problem with general properties of interacting many-body systems. Finally, we provide a concise review of the 
rigorous results on equilibration times that are known in the literature.  
\end{abstract}

\section{Introduction} \label{sec:introduction}

The observation that closed quantum systems with many degrees of freedom generically equilibrate to a seemingly
stationary state has already intrigued the forefathers of quantum mechanics \cite{Schroedinger1927,vonNeumann1929}. Indeed,
such complex quantum systems seemingly relax to stationarity, despite the entire system undergoing perfectly
unitary dynamics. This is not a contradiction: Unitary dynamics is compatible with many observables relaxing in their
expectation values to high accuracy, such that the coherent time-evolution can only be witnessed by measuring complex, global observables to high accuracy. 

Insights into equilibration of quantum many-body systems are at the heart of the 
foundations of quantum statistical mechanics:
After all, the notion of an equilibrium ensemble naturally makes sense only if one can think of stationary properties,
and if these are to be compatible with the microscopic laws of quantum mechanics, they have to emerge
from quantum dynamics in one way or the other (see the reviews Refs. \cite{Gogolin2016,Polkovnikov2011} and, e.g., 
Refs.\ \cite{Cramer2008,Linden2009,Reimann2008,Short2012,Reimann2012a,Farrelly2016}). 
These conceptual considerations are largely backed up by a body of numerical
studies (see again the reviews Refs.\ \cite{Polkovnikov2011,EisFriGog15} and, e.g., Refs.\ \cite{Rigol2007,PhysRevA.80.053607,NumericalQuench2,PhysRevE.88.032913,PhysRevB.85.085129,PhysRevB.81.115131,PhysRevLett.112.065301}
for a selection of works). 
Indeed, much of the interest in the question of equilibration is motivated by and stems from  
research questions on the foundations of statistical mechanics. 

More recently, and equally importantly, questions of equilibration have risen to 
prominence again due to the fact that they feature strongly in the analysis of quenched quantum many-body
systems out of equilibrium \cite{EisFriGog15,Polkovnikov2011}, in a way they can be precisely probed and explored with cold atomic systems \cite{Bloch2012} and other controlled quantum systems including trapped ions \cite{BlattSimulator}. 
Since the time-evolution of complex many-body systems cannot be efficiently simulated using classical computers, but may be simulated in such experimental set-ups, many of the recent efforts of dynamical quantum simulation also hint at
or build upon questions of equilibration. 
While the basic mechanism of equilibration of
closed quantum systems due to dephasing is largely understood, much 
less is known on the times at which this is expected to happen. Indeed, since much of the question of equilibration times
is still open, quantum simulation allows us to assess a regime of quantum many-body physics that can not yet be backed up in all details by a theoretical underpinning. 

This book chapter addresses the questions 
of what is known on equilibration times of closed quantum many-body systems. 
We aim at both providing an intuitive understanding of equilibration in terms of dephasing in connection with physically plausible assumptions on quantum many-body system as well as providing a concise review of the rigorous results available in the literature, in the hope to motivate more researchers to work on this interesting and interdisciplinary problem.



\section{Quench experiments and non-equilibrium dynamics}
Before coming to the theoretical discussion of equilibration, let us briefly discuss how equilibration 
of complex many-body systems can be studied experimentally.
One of the most prominent architectures for probing out of equilibrium dynamics of quantum many-body systems
is constituted by cold atoms in optical lattices or in the continuum, another is that of trapped ions.
In fact, some of the research questions addressed in this book
chapter have been triggered by experimental findings from that context that have not yet found a 
satisfactory explanation.

\subsection{Cold atomic settings}

One of the earliest experiments with cold atoms in optical lattices was concerned with a sudden ``quench''
in which the Hamiltonian parameters were rapidly changed from a superfluid to a Mott phase 
and the subsequent non-equilibrium dynamics monitored \cite{Collapse}. Genuine equilibration was
observed in a setting in which a charge density wave was initially prepared, making use of an optical
superlattice, quenched to an interacting many-body Hamiltonian well captured by a Bose-Hubbard model
\cite{Trotzky2012}. In such a setting, several quantities can then be precisely observed as they
evolve in time, prominently the imbalance \cite{NumericalQuench2} between 
odd and even sites of the lattice. This quantity exhibits a characteristic equilibration dynamics,  following a power
law in time in the close to integrable settings \cite{Cramer2008,NumericalQuench2}. Since then, several settings featuring equilibration have been studied 
\cite{EisFriGog15,Polkovnikov2011,Bloch2012,Kaufman,Schreiber2015}. 
Importantly, systems featuring many-body localisation \cite{Schreiber2015} equilibrate
in the sense that the state becomes locally practically indistinguishable from its time average for most
times. However, they do not 
thermalise, in that the expectation values obtained are different from the ones of the canonical 
ensemble.
Ref.\ \cite{Kaufman} observes specifically  local equilibration and thermalisation, while showing the
coherence of the full evolution. In continuous systems of cold atoms \cite{Gring2012}, 
similar features of equilibration are observed. 

\subsection{Trapped ions and hot electrons}
Complementing this development, systems of trapped ions \cite{BlattSimulator} 
allow us to monitor equilibration dynamics in time. For example, Refs. \cite{PhysRevLett.119.080501,53}
observe dynamical quantum phase transitions, but along the way also notice 
features of equilibration. Another example is 
the experimental realisation of a physical system featuring many-body localisation in a
system of trapped ions with programmable disorder
\cite{MonroeMBL}, again exhibiting equilibration in time.
Having said that, experimental studies of non-equilibrium
dynamics in the sense discussed here is by no means confined to cold atomic systems or
settings of trapped ions: Ref.\ \cite{HotElectrons}, e.g., shows ultra-fast relaxation of hot electrons.
In all these setting, questions of equilibration times arise, further motivating the endeavors described in this chapter.

\section{Heuristic discussion of equilibration}
\subsection{What it means for a closed system to equilibrate}
%
%
In this section, we aim to establish an intuitive understanding of how equilibration in closed systems happens and why it seems plausible that it happens quickly in a generic many-body system.
The discussion in this section follows Refs.~\cite{DeOliveira2017,Wilming2017c}, where more detailed expositions can be found. 
In section~\ref{sec:rigorous}, we then present rigorous results on equilibration, both in finite and in infinite time.

Consider a finite quantum system, described by Hamiltonian $H$ with spectral decomposition
\begin{equation}
H=\sum_{k=1}^{d_E} E_k P_k.
\end{equation}
Here, $E_k$ are the eigenvalues and $P_k$ the projectors onto its eigenspaces,
which can be degenerate so that $d_E$ may be smaller than the
total Hilbert space dimension $d_T$.
As we are interested in studying the dynamics of a closed quantum
systems, we will assume that the system is initially in a pure state vector, which can be written as $|\psi_0 \rangle =\sum_k c_k |E_k\rangle$
with $c_k=\langle E_k | \psi_0\rangle$. We can always choose a basis in each degenerate energy-eigenspace $P_k$ so that $|\psi_0\rangle$ only has overlap with one basis-vector $|E_k\rangle$ in this subspace and in the following always assume this choice of basis.
As time evolves, the state vector of
the system is given by\footnote{Note that we take $\hbar=1$ throughout.}
\begin{equation}
|\psi(t) \rangle = \sum c_k \; e^{-i E_k t} \;|E_k\rangle .
\end{equation}
Unless the initial state is an eigenstate of $H$, the system will
never stop evolving and in this sense the system never equilibrates.
But for quantum many-body systems we also do not expect to have access to the
instantaneous full quantum state of the system as we would need to keep track of an astronomical number of observables.
Usually, we are only interested in a small, fixed set of observables, such as local observables. Also interesting
and physically plausible are often sums of local terms, such as the magnetisation in a spin-system. 
At this level we then may have equilibration due to the fact that we are not accessing all the information about the system.

Let us assume we are interested in some observable $A$, whose expectation value
evolves in time as 
\begin{equation}
	\langle \psi(t)|A|\psi(t) \rangle = \sum_{i,j} c_j^* \; A_{j,i} \; c_i \; e^{-i(E_{i}-E_{j})t}
\end{equation}
with $A_{i,j} = \langle E_i | A | E_j \rangle$. The question is then if
such observable can equilibrate. A system that is equilibrating has to 
equilibrate to the infinite time-average
\begin{equation}
	\lim_{T\rightarrow \infty} \frac{1}{T} \int_0^T \langle \psi(t) | A|\psi(t) \rangle \mathrm{d}t = \sum_i |c_i|^2 A_{i,i},
\end{equation} 
since the expectation value of $A$ for an equilibrating system is close to a particular value for the vast majority of the time and hence the time-average of $A$ is also close to this value \footnote{Furthermore, it is  straightforward to show that $\lim_{T\rightarrow \infty} \frac{1}{T} \int_0^T \left( \langle \psi(t) | A | \psi(t) \rangle - A_{\mathrm{eq}}\right)^2 \mathrm{d}t $ is minimised by setting $A_{\mathrm{eq}} =\lim_{T\rightarrow \infty} \frac{1}{T} \int_0^T \langle \psi(t)|A|\psi(t) \rangle \mathrm{d}t$.}.
The time-averaged expectation value corresponds to the expectation value
of $A$ in the quantum state $\omega=\sum |c_i|^2 |E_i\rangle\langle E_i|$ that maximises the von~Neumann entropy given all the conserved quantities
of the dynamics \cite{Gogolin2011}. However, in any finite system there will be recurrences, so that $\langle \psi(t)|A|\psi(t)\rangle$ is in fact quasi-periodic and never equilibrates perfectly \cite{Wallace2013}. 
Nevertheless, it can happen, and indeed often does happen, that the deviation of $\langle \psi(t)|A|\psi(t) \rangle$ from its time-average, which is given by
\begin{eqnarray}
\Delta A(t) & = \sum_{i \neq j} c_j^* \; A_{j,i} \; c_i \; e^{-i(E_{i}-E_{j})t},
\end{eqnarray}
is undetectably small for most of the time. 
To show this, one often analyses the infinite time-average of the fluctuations $\Delta A(t)^2$ and we will later present rigorous results which show that this quantity is often extremely small in large systems. If this is the case, i.e., if the time-average of the fluctuations is very small, then the fluctuation $\Delta A(t)^2$ is small for most times. 
 It is in this sense, that typically the fluctuations are undetectable, that we
can meaningfully speak about equilibration of closed quantum systems. 

It is important to stress that this does not say much about how long it takes the system to reach equilibrium.
If a system of $N\sim 10^{23}$ particles takes a time that is exponential in $N$ to reach equilibrium, it practically does not equilibrate.
Very roughly speaking, we will thus say that a system equilibrates quickly if the time it takes to reach equilibrium does not depend strongly on the physical size of the system. To explain equilibration in many-body systems, it is thus necessary to explain both why such systems equilibrate at all and why the time it takes them to equilibrate does not increase strongly with the system size. 

So far we have talked only about the equilibration of the expectation value of a given observable. 
A more stringent notion of equilibration requires that the whole probability distribution of measurement outcomes of a given observable equilibrates. If $A=\sum_{\lambda=1}^{d_A} a_\lambda P_\lambda$ is the spectral decomposition of $A$, we thus require that all the spectral projections $P_\lambda$ equilibrate in expectation value. In the following heuristic discussions, we do not emphasise this but simply assume that the arguments also apply for the projectors $P_\lambda$, but this point will be discussed more thoroughly in the section on rigorous results.
For now, simply note that in many-body systems the physical relevant observables are usually local observables and the corresponding projectors $P_\lambda$ are also local observables. 
Thus, any argument that shows equilibration for all local observables also shows equilibration of their measurement statistics, justifying this simplification. 

\subsection{Intuitive understanding of equilibration as dephasing}
\label{sec:intuitive}
Having explained in which sense we can say that a closed, finite, quantum system equilibrates, let us now start to develop an intuitive explanation of this process.
The expression for the fluctuations away from equilibrium can be rewritten as 
\begin{eqnarray}\label{eq:deltaA}
	\Delta A(t) = \sum_\alpha v_\alpha e^{-i G_\alpha t},
\label{eq:fluct}
\end{eqnarray}
where $\alpha \in \mathcal{G}=\{(i,j):i,j\in\{1,\ldots d_{E}\},i\ne j\}$ labels the energy gaps $G_{\alpha}=(E_{j}-E_{i})$ appearing in the system's spectrum
and $v_\alpha= c_j^* \; A_{j,i} \; c_i $. The expression
is similar to a Fourier series, but it is not a Fourier series as in general $G_\alpha$ are not
multiples of some fundamental ``frequency" and we hence obtain a quasi-periodic
function, whose rigorous mathematical treatment is complex in general.

\begin{figure}[t]
	\includegraphics[width=15cm]{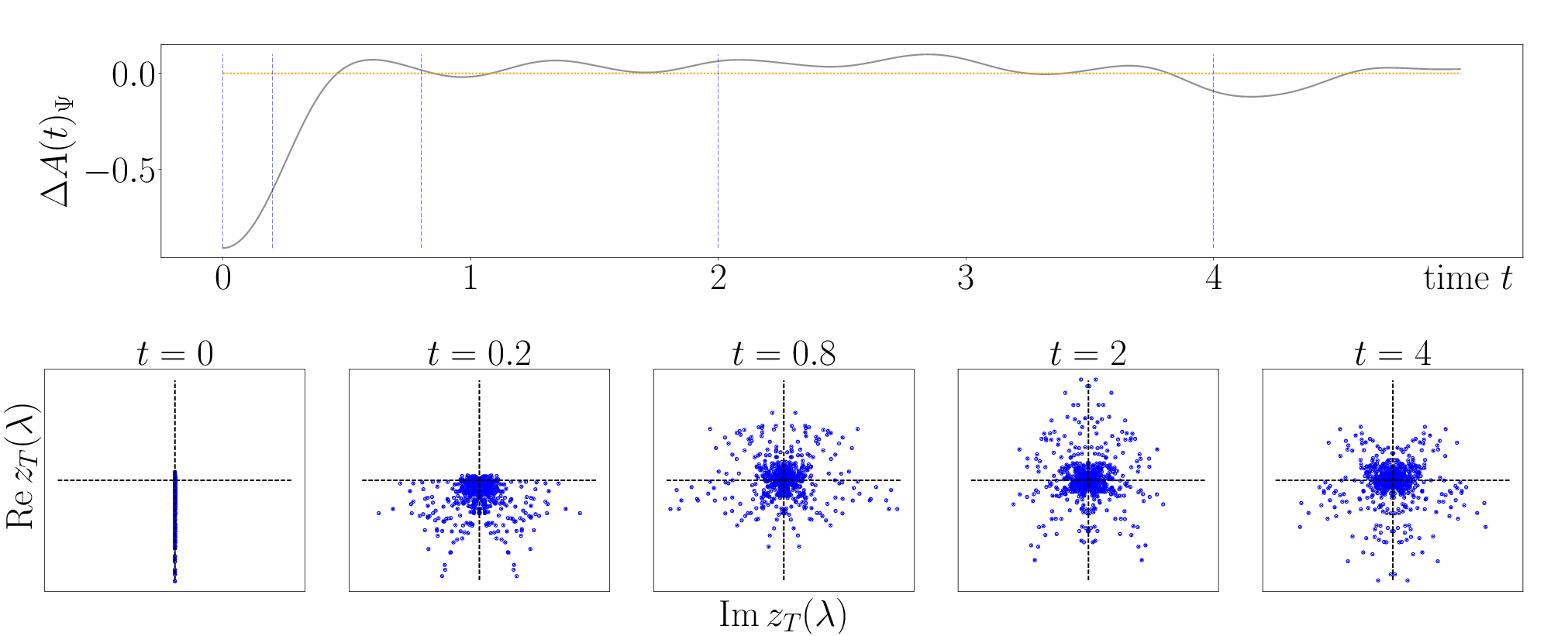}
	\caption{Time-evolution of a XXZ model with next-nearest neighbour interaction on 15 lattice sites and charge-density wave as initial state $\Psi$ (see Ref.~\cite{Wilming2017c} for details). 
		The upper panel shows the time-evolution of $\Delta A(t)$, where the observable is given by $\sigma_z$ on a single spin. 
	The lower panels show the time-evolution of the regularised $z_G$ in the complex plane (see paragraph before \eqref{eq:limit}), which are here denoted by $z_T(\lambda)$ instead of $z_T^N(G)$, for $T\approx 33$. (Figure from Ref.~\cite{Wilming2017c}.)}
	\label{fig:evolutionXXZ}
\end{figure}

Thus we will in the following first try to give intuitive and heuristic arguments about its
behaviour and therefore about equilibration. One way to understand
the behaviour of $\Delta A(t)$ is to consider each term $v_\alpha e^{-i G_\alpha t}$ as a vector or point in the complex plane evolving with time, with each point moving on a circle of radius $|v_\alpha|$ with angular velocity $G_\alpha$. 
We can thus think about these points as a cloud of points evolving in time, see Fig.~\ref{fig:evolutionXXZ} for a numerical example using a XXZ model and Ref.~\cite{Wilming2017c} for further numerical examples of systems that do equilibrate and systems that do not equilibrate.

As the value of $\Delta A(t)$ is the total vector, to have a large
fluctuation we need most of the vectors pointing roughly in the same
direction, i.e., the cloud of points cannot be isotropic. 
This suggests that in a given initial state, randomly chosen observables are typically already equilibrated (also see section~\ref{sec:rigorous} and the chapter by Balz et al.\ in this book). 
Suppose now that the initial state is out of equilibrium: $\Delta A(t)$ is large and most of the vectors $v_\alpha$ point
in the same direction. 
As time evolves each vector $v_\alpha$ will
start to rotate with a angular velocity $G_\alpha$. 
If we assume that every gap $G_\alpha$ is unique, all the points move with a different velocity and therefore
the vectors will start to distribute more isotropically in the complex plane and their sum will become small. 
In the case where we have $G_\alpha=G_\beta$ for $\alpha\neq \beta$, we can first regroup the vectors into new vectors $z_G=\sum v_\alpha$, where the sum is over all $\alpha$ with $G_\alpha=G$ and then apply the same reasoning to the representation of $\Delta A(t)$ as 
\begin{align}
	\Delta A(t) = \sum_{G\in\mathsf{Gaps}} z_G \e^{- i G t},
\end{align}
where the set $\mathsf{Gaps}=\{G_\alpha : \alpha\in \mathcal G\}$ is the set of different energy gaps. 
This mechanism, usually called dephasing, occurs in many physical phenomena, as the spreading of a wave propagating in a dispersive media or the spreading of the wave-function of a particle in quantum mechanics. 
The essential difference to our case is that we have a discrete distribution of points and frequencies and not a smooth distribution of points and frequencies, which makes the analysis more complex. 

There are hence two fundamental ingredients here: i) the number and distribution of the vectors $v_\alpha$
contributing to the sum, ii) the distribution of the values of $G_\alpha$. 
If there are just a few vectors, or if most of them are negligible, then the
vectors will align again in a short time: we will have oscillations and not
equilibration. On the other hand if there are many of them it will take a long
time to have a realignment; in fact this is the recurrence time which typically
increases very fast with the number of vectors. Besides, it can be shown that
the typical value of the fluctuation is upper bounded by $\sum_\alpha |v_\alpha|^2$.
Therefore to have equilibration we need many $v_\alpha$ contributing to the sum.
This is typically the case for generic initial states in many-body systems, since there are roughly $d_E^2$ energy gaps in the spectrum, which is a number exponentially large in the system size, and generic states will have small overlap with all energy eigenvectors.

Regarding the angular velocities, $G_\alpha$, to have equilibration in short time, they must not have a sharp distribution, since then
the time for the vectors to spread will be very large since they only disperse very slowly. 
 In sum, to have good equilibration we need many vectors contributing and the time for it to happen depends on the distribution of the values of $G_\alpha$ and the corresponding amplitudes $v_\alpha$. 
 But how do the equilibration properties of the system depend on the distributions of these quantities?

One way to approach the problem is to assume that the distributions of $v_\alpha$ and
$G_\alpha$ can be well approximated by a smooth distribution. By this, we mean that
\begin{align}
	\Delta A(t) &= \sum_\alpha v_\alpha e^{-i G_\alpha t} = \int \mu(G) v(G) e^{-i G t} \mathrm{d} G \\& \approx \int z(G) \e^{-iGt} \mathrm{d} G,\label{eq:approximationbysmoothfunction}
\end{align}
where $v$ is the distribution of the $v_\alpha$ and $\mu$ is the density of energy-gaps (in a distributional sense) and we assume that $z$ is a continuous function. 
The most important case where we can hope to make sense of such an approximation is that of a scaling limit of a many-body system, in which the recurrent nature of the dynamics of the finite dimensional quantum system is broken. 
Suppose therefore, that we have a sequence of system-sizes $N$ and choose for every system-size a Hamiltonian, observable and initial state in a compatible manner.
The prototypical example is given by a translational invariant local Hamiltonian on a square lattice with a translational invariant, pure product state as initial state (possibly with a larger periodicity in space than the Hamiltonian, such as a charge density wave) and with $A$ being a fixed local observable around the origin of the lattice.
In this case, Lieb-Robinson bounds \cite{Lieb1972,Kliesch2014} imply that if the system can be shown to equilibrate in the thermodynamic limit $N\rightarrow \infty$ in time $\tau$, then it will also equilibrate in time $\tau$ and remain equilibrated for a long time for sufficiently large, but finite systems. 

We can thus hope that a continuous distribution $z$ emerges in the thermodynamic limit and use the form of this function to argue about the equilibration time. 
To be more precise about how such convergence may be understood we should first regularize the discrete distribution of $z_G$ (or $v_\alpha$) into a smooth distribution, for example by convolving with a Gaussian of variance $1/T$, yielding a smooth distribution $z^N_T$ for every system size $N$ of the sequence of initial states, Hamiltonians and observables that we consider. 
Eq. \eqref{eq:approximationbysmoothfunction} then holds in terms of $z^N_T$ for times much smaller than $T$. 
The statement that the distribution of $v_\alpha$ and $G_\alpha$ converge to a continuous function then means that the limit 
\begin{align}\label{eq:limit}
	\lim_{T\rightarrow \infty} \lim_{N\rightarrow\infty} z^{N}_T(G) = z(G)
\end{align}
exists and yields a continuous function. 

The fluctuations in time are then approximated by the Fourier transform of the function
$z$. As a simple example, assume that $z$ is real and given by a Gaussian with mean zero and variance $1/\tau$. We then 
have
\begin{equation}
\Delta A(t) \approx \Delta A(0) \; e^{(-t/\tau)^2}.
\end{equation}
Thus $\Delta A(t)$ is a also a Gaussian with mean zero and variance $\tau$. 
Using the variance as the scale for the decay of the Gaussian we can identify $\tau$ as an equilibration time-scale: the equilibration time is the variance of $\Delta A(t)$, which is proportional to the inverse of the variance of $z$. 
A similar relationship, which is a kind of uncertainty principle, can be expected to hold whenever $z$ is a bounded, (square-)integrable function and is roughly unimodal, i.e., only has one strong peak. 
In particular, whenever $z$ is square-integrable, $\Delta A(t)$ is also square-integrable and hence has to decay to zero as $|t|\rightarrow \infty$.
We will later provide arguments that make this form of $z$ plausible for generic many-body systems if the observable $A$ is a local observable. 
The rigorous results presented in section~\ref{sec:rigorous} will further elaborate on the connection between the equilibration time and the form of the probability distribution $p_\alpha=|v_\alpha| / Q$, where $Q = \sum_\alpha |v_\alpha|$.

\subsection{Connecting to general properties of many-body systems}
As emphasized in the previous section, the equilibration behaviour depends on the initial state, Hamiltonian and the observable one is interested in. 
In this section we specialise to the case of a local many-body system and present general properties of such systems that 
add plausibility to the assumptions that we made in the previous section. In the following we thus consider a situation,
where i) the Hamiltonian is a local Hamiltonian 
$H = \sum_{x\in\Lambda} h_x$ on some regular lattice $\Lambda$ with $|\Lambda|=N$ lattice sites; 
ii) the observable $A$ is a local observable, i.e., supported on some finite region independent of the system size, for example a spin in the center of the system; 
iii) the initial state $\rho$ is pure and has a finite correlation length $\xi>0$,
\begin{align}
	\tr(\rho A B)-\tr(\rho A)\tr(\rho B) \leq \norm{A}\norm{B} \e^{- d(A,B)/\xi},
\end{align}
where $d(A,B)$ is the lattice distance between the support of the observables $A$ and $B$ and $\|.\|$ denotes the
operator norm, so the largest singular value.
The last assumption is of particular relevance for quench experiments in optical lattices, where the initial state is often given
by either a well controlled product state or the ground-state of some non-critical, local Hamiltonian.

Let us begin with discussing the relevant properties of the underlying Hamiltonian. 
In a quench experiment, the initial state $\rho$ can usually be expected to have a finite energy density with respect to the Hamiltonian $H$. 
It will thus be supported not on the low-energy subspace close to the groundstate, but in the bulk of the spectrum. 
We are hence interested in how the bulk of the energy-spectrum looks like for generic, local Hamiltonians. 
The first important observation to make is that the number of different eigenvalues of a local Hamiltonian is typically exponentially large in the size of the system, while their magnitude is at most linear in the size of the system. 
Therefore, at least in the bulk of the spectrum, the spectrum is extremely dense and typical differences between neighbouring eigenvalues are exponentially small in the system size. 
Indeed, it is well known that the energy-spectrum of a generic local Hamiltonian can be well approximated by a Gaussian in the bulk of the spectrum.
To understand this, observe that the energy spectrum can be seen as the probability distribution of energy in the maximally mixed state $\one/d_T$. 
Since this state is a product-state and the Hamiltonian consists of a large sum of operators, with the support of each of them overlapping only with the support of finitely many other ones, we can understand the spectrum of the Hamiltonian as a large sum of weakly-correlated, bounded random variables. 
We can hence expect that a central-limit theorem applies, yielding a Gaussian density of states in the bulk of the spectrum.
Indeed, such arguments can be made rigorous, showing that for any state with a finite correlation length, the distribution of energies $p_i = \tr(\rho |E_i\rangle\langle E_i|)$ is roughly Gaussian with a standard deviation of order $\sqrt{N}$ \cite{BrandaoCramer2015,Keating2015,Anshu2016}. 
As a consequence of these results, we can also expect that the distribution of energy gaps follows a roughly Gaussian distribution with standard deviation of order $\sqrt{N}$. 
Since energy gaps always come in pairs $G_{(i,j)} = - G_{(j,i)}$, this distribution has mean zero. 
At this point, it is worth emphasizing that we are here talking about the full distribution of energy-differences in the spectrum of the many-body Hamiltonian, and \emph{not} about what is known as the \emph{level-statistics} in random matrix theory (see chapter by Santos and Torres-Herrera in this book), which is concerend with the expected distance in energy between the $i$-th energy level and the $i+1$-th energy level (or, more generally, the $i+k$-th energy level).

The above discussion already suggests that for a pure initial state with finite correlation length and finite energy density, the expansion coefficients $c_i$ in the energy eigenbasis can be thought of as a smooth distribution that is spread out over exponentially many energy-levels, with each $c_i$ being exponentially small in absolute value. 
Indeed, there are several arguments supporting that one can expect that the \emph{inverse participation ratio} $\mathrm{IPR}$ or the inverse of the \emph{effective dimension} $\deff$ are exponentially small in the system size for generic, interacting many-body systems \cite{Linden2009,Gogolin2011,Reimann2012,Reimann2012a,Farrelly2016,Gallego2017,Wilming2018},
\begin{align}\label{eq:deffIPR}
	\sum_i |c_i|^4 = \mathrm{IPR}(\rho,H) = \frac{1}{\deff(\rho,H)} \leq \e^{- k N}, 
\end{align}
for some constant $k>0$. 
This quantity will also play an important role in the following section, treating rigorous results about equilibration.

Let us now turn to the observable $A$. Intuitively, a local observable should only be able to connect energy eigenstates which differ by a small amount, suggesting that matrix elements $A_{i,j}$ are very small if $|E_i-E_j|$ is large. This can indeed be made rigorous, as
has been shown in Refs.~\cite{Arad2014,DeOliveira2017}. In a local many-body system and for any fixed local observable there exists constants $\alpha,R>0$ such that
\begin{align}
	|A_{i,j}| \leq \norm{A} \e^{-\alpha(|E_i-E_j| - 2R)},
\end{align}
where $R$ is proportional to the support of $A$. This implies that the coefficients $v_\alpha$ are exponentially small in $G_\alpha$ and the function $z$ can be expected to fall off exponentially in $|G|$.
Since the gaps are distributed essentially like a Gaussian with standard deviation of order $\sqrt{N}$, this implies that the on the scale of the gaps that are relevant to the problem, the distribution of gaps can be expected to be essentially \emph{uniform}. 
Thus we can, by making only a small error, replace the distribution $\mu$ in \eqref{eq:approximationbysmoothfunction} by $1/2G_{\mathrm{max}}$, where $G_{\mathrm{max}}$ is some cut-off gap.
As long as the distribution $v$ is a well-defined bounded function, we then obtain equilibration in a time that does not diverge with the system size, since the function $z$ will be bounded and integrable. 
Since we expect the coefficients $c_i$ and hence the $v_\alpha$ to be exponentially small in $N$, this seems highly plausible. 
However, since the number of coefficients $v_\alpha$ is also exponentially large in $N$, it is in principle possible that as we increase the system size, exponentially many of them concentrate in an exponentially small region of gaps $G$, leading to a situation where $v$ (and hence $z$) is not given by a bounded function, but can only be understood in a distributional sense.    
Thus, while the above arguments make it plausible that generical many-body systems equilibrate quickly and allow us to understand in a qualitative way how this happends, they do not provide a riogorous proof. 
Having discussed the heuristics of equilibration in many-body systems, let us now turn to rigorous, general results about equilibration in closed quantum systems.

\section{Rigorous results}
\label{sec:rigorous}

Consider an arbitrary initial state $\rho(0)$ (which may be pure or mixed) of a finite-dimensional quantum system, evolving via a Hamiltonian $H=\sum_{k=1}^{d_E} E_k P_k$. Let us denote the time averaging of an arbitrary quantity $f(.)$ over a finite interval of time $T$ by 
\begin{equation} 
\av{f(t)}_T= \frac{1}{T} \int_0^T f(t) \, dt,
\end{equation}
where $\av{f(t)}_\infty = \lim_{T \rightarrow \infty} \av{f(t)}_T$. 
For the state to equilibrate with respect to a given observable $A$  it is necessary for the expectation value of $A$ for $\rho(t)$ to be very close to the expectation value of $A$ for the static equilibrium state $\omega = \av{\rho(t)}_\infty$ for most times. This happens under very general conditions. Indeed, all that is required is that the state is spread over many different energies, and that the Hamiltonian does not contain any highly degenerate energy gaps in its spectrum. In particular, it can be proven that \cite{Reimann2008, Short2012} 
\begin{equation} 
\av{(\tr(\rho(t) A) -  \tr(\omega A))^2 }_\infty \leq g \sum_\alpha |v_\alpha|^2  \leq \frac{g\| A\|^2}{\deff}, \label{eq:equil1}
\end{equation} 
where $g$ is the degeneracy of the most degenerate energy gap\footnote{I.e. $g=\max_\alpha | \{\beta :  G_{\beta} = G_{\alpha}\}|$.}, $v_\alpha= \rho_{i,j} A_{j,i}$ is the quantity described in section~\ref{sec:intuitive} (generalised slightly to include initial mixed states), $\|.\|$ again
denotes  the operator norm, and 
\begin{equation} 
\deff = \frac{1}{\sum_k \tr(P_k \rho(0))^2},
\end{equation} 
is the \emph{effective dimension} of the state, describing approximately how many different energies it is spread over (e.g. if the state is spread equally over $N$ different energy levels then $\deff=N$). Hence if $g$ is not too large, as one would expect for a physically realistic interacting Hamiltonian, and $\deff$  is large, as one would expect for a realistic quantum many-body system state, then $\av{(\tr(\rho(t) A) -  \tr(\omega A))^2 }_\infty \ll  \| A\|^2$ and the expectation value of $A$ equilibrates. One expects the effective dimension to grow with the system size, leading to less pronounced
deviations from the time average. Again, there is strong numerical evidence for this expectation \cite{PhysRevE.88.032913} and theoretical arguments which suggest that $\deff$ grows exponentially with $N$ in many interacting many-body systems, see references before \eqref{eq:deffIPR}.

Note that although equilibration of the expectation value is a necessary condition for equilibration, it is not by itself sufficient, as one can construct very different observable distributions with the same expectation value. 
A stronger definition of equilibration with respect to an observable $A$, is to show that for most times one cannot distinguish $\rho(t)$ from $\omega$ via a measurement of $A$. 
We define the distinguishability $D_A( \rho(t), \omega)$ as the statistical distance between the probability distributions obtained when measuring $A$ on $\rho(t)$ and $\omega$ \footnote{Hence $D_A( \rho(t), \omega) = \frac{1}{2} \sum_i |p_i(\rho(t)) - p_i(\omega)|$, where $p_i(\rho(t))$ is the probability for result $i$ in a measurement of $A$ on state $\rho(t)$. 
This can  be understood operationally in terms of the maximal success probability $p$ of guessing correctly whether the state is $\rho(t)$ or $\omega$ after measuring $A$ (given that you are given either  $\rho(t)$ or $\omega$ with equal probability) via $D_A( \rho(t), \omega) = 2p-1$.}. The previous result can then be used to obtain a bound on the distinguishability \cite{Short2011, Short2012,Malabarba2016}, giving
\begin{equation} 
\av{D_A( \rho(t), \omega)}_\infty \leq \frac{1}{2} \left({ \frac{g (N-1)}{\deff}}\right)^{1/2},\label{eq:equil2}
\end{equation} 
where $N$ is the number of possible outcomes in the measurement of $A$ (i.e., the number of distinct eigenvalues of $A$), which is typically much less than $\deff$ for realistic measurements on quantum many-body systems. Similar results can  be obtained for finite sets of measurements \cite{Short2011, Short2012}, or for all possible measurements on a small subsystem, proving that small subsystems interacting with a large bath generally equilibrate to a static reduced density operator \cite{Linden2009,Short2012}.

The above results apply to infinite time equilibration, but can be extended to equilibration over a finite time interval \cite{Short2012}. In the case of \eqref{eq:equil1} this gives
\begin{equation} 
\av{(\tr(\rho(t) A) -  \tr(\omega A))^2 }_T \leq \frac{g \| A\|^2}{\deff} \left( 1 + \frac{8 \log_2 d_E}{\epsilon_{\min} T} \right),
\end{equation} 
where $\epsilon_{\min}$ is the smallest difference between energy gaps\footnote{If desired, one can replace $\epsilon_{\min}$ with an arbitrary energy $\epsilon>0$, and $g$ by $N(\epsilon)$, the maximum number of energy gaps which fit within a window of size $\epsilon>0$.} (i.e., $\epsilon_{\min} = \min_{\alpha \neq \beta} |G_{\alpha} - G_{\beta}|$) . The $\log_2 d_E$ term is slightly awkward, as it means that the bound does not extend to infinite dimensional systems with discrete spectra. However, a different approach \cite{Malabarba2014} can eliminate this term at the expense of a slightly worse infinite-time limit, giving 
\begin{equation} 
\av{(\tr(\rho(t) A) -  \tr(\omega A))^2 }_T \leq \frac{g \| A\|^2}{\deff} \frac{5 \pi}{2}  \left( \frac{3}{2} + \frac{1}{\epsilon_{\min} T} \right).
\end{equation}
Similarly to before, this result can also be used to bound the distinguishability via
\begin{equation} 
\av{D_A( \rho(t), \omega)}_T \leq \frac{1}{2}  \left(  \frac{g (N-1)}{\deff} \frac{5 \pi}{2} \left( \frac{3}{2} + \frac{1}{\epsilon_{\min} T} \right) \right)^{1/2}.
\end{equation} 
These results are very general, but as a consequence they generally lead to very large equilibration time bounds. In particular, consider a system whose state is prepared in an energy window of width $\Delta E$ containing $d$ states. Then even in the best case we would have $\deff = d$ and $\epsilon_{\rm min} \approx {2\Delta E}/{d^2}$ (as there are $d^2$ energy gaps between $d$ levels, and the range of gaps is twice as large as the energy range), requiring $T \approx {d}/{\Delta E}$ for the bound to become significant. This is much shorter than the recurrence time (which is typically exponential in the dimension \cite{Hemmer1958}) but is still much larger than observed equilibration times for realistic physical systems. 

One might wonder whether this general bound could be tightened significantly, or whether systems could exist which really required such large equilibration times. The answer is the latter \cite{Goldstein2013, Malabarba2014}. Indeed for any initial pure state with high effective dimension,  we can construct an observable which takes an extremely long time to equilibrate. Consider the projector onto the subspace spanned by  `snapshots' of the evolving state for many successive discrete time steps. By choosing an appropriate size and number of time-steps, one can show that this observable will take longer than $ {\deff}/{(1000 \sigma_E)}$ to equilibrate \cite{Malabarba2014} (where $\sigma_E$ is the standard deviation in energy). This has a similar scaling to the general bound considered above. 

One limited situation in which fast equilibration can be proven is when the observable to be measured has only two possible outcomes, and the rank of the projector onto one of the outcomes (which we will denote by $K$) is very small. In this case, one can show that \cite{Malabarba2014}
\begin{equation} \label{eq:smallK}
\av{D_A( \rho(t), \omega)}_T \leq c \left({ K \eta_{\frac{1}{T}}}\right)^{1/2},
\end{equation} 
where $c \simeq 7$ and 
$\eta_{\frac{1}{T}}$ is the maximum probability of the state lying in an energy window of width ${1}/{T}$ (i.e., the maximum over $E$ of the probability of the energy being in the range $(E, E+{1}/{T})$). For an initial state with a dense set of occupied energy levels, we could approximate the energy distribution via a continuous function. If this function is approximately unimodal (i.e., with `one hump', such as a Gaussian or top-hat function) then the maximum probability density will be 
$\sim {1}/{\sigma_E}$. In such a case, we would obtain  
$\eta_{\frac{1}{T}} \sim {1}/({\sigma_E T })$. More generally, we can always define constants $a$ and $\delta$ such that 
\begin{equation} \label{eq:etabound}
\eta_{\frac{1}{T}} \leq \frac{a}{\sigma_E T } + \delta,
\end{equation} 
where $a>0$ is a real parameter which captures the shape of the distribution and $\delta>0$ corrects for the discreteness of the spectrum. For approximately unimodal energy distributions spread over many energy levels, we would expect $a\sim 1$ and $\delta \ll 1$. Inserting (\ref{eq:etabound}) into (\ref{eq:smallK}) we obtain   
\begin{equation} 
\av{D_A( \rho(t), \omega)}_T \leq c \left({ K \left(\frac{a}{\sigma_E T } + \delta \right)}\right)^{1/2},
\end{equation} 
giving good equilibration after $T \sim{1000 a K}/{\sigma_E}$, a typically fast time scale with no explicit dependence on $\deff$. When the initial state is pure, a particularly interesting case of such an observable  is the projection onto the initial state, for which $K=1$. We will return to this example in the next subsection. 

Another possibility is to calculate equilibrium times for specific systems. In Ref.\ \cite{Cramer2008} 
(see also Refs.\ \cite{Gluza2016,Calabrese2012}) it is shown that the Bose-Hubbard model quenched from a Mott quantum phase to the free strong superfluid regime obeys local equilibration over the entire interval $[t_{\mathrm{Relax}}, t_{\mathrm{Relax}} + t_{\mathrm{Recurrence}} ]$ (i.e.,  individual sites or small blocks are almost indistinguishable from a static state for all times in this interval). The equilibration time,  $t_{\mathrm{Relax}}$, is relatively fast, and is governed by the inverse of the hopping parameter (which determines the speed of sound in the system) and the desired equilibration closeness, whilst $t_{\mathrm{Recurrence}}$ can be made arbitrarily large by increasing the size of the system. 

\subsection{Bounding equilibration times using randomness} 

Although we have shown that in general equilibration times can be very large, most observables of interest in the real world seem to equilibrate much faster -- typically in time scales which depend on the physical size of the system rather than its dimension (which for a many body system is given by $\log d$ rather than $d$).   An interesting question is  to consider the equilibration times of `typical' situations, which might equilibrate much faster than the general bound. One way to approach this is to choose one of the components of the setup (the observable, Hamiltonian, or initial state) at random - thereby avoiding fine-tuned  setups. The effect of introducing randomness in each of the three components has been considered \cite{Masanes2013, Malabarba2014,  Reimann2016, Garcia-Pintos2015a, Balz2017,Brandao2012,Cramer2012,Goldstein2015, Vinayak, Masanes2013}, and does indeed lead to much faster equilibration times.

\subsubsection{Random observables} Let us first consider the equilibration of a randomly chosen observable, given a fixed Hamiltonian and a fixed pure initial state with high effective dimension. Note firstly that most observables are already equilibrated (assuming they have a reasonable number of distinct outcomes $N \ll d$), in the sense that they cannot distinguish the true state $\rho(t)$ from the equilibrium state $\omega$ over any interval. In particular
\begin{equation} 
\av{D_A( \rho(t), \omega)}_A \leq \frac{1}{2} \left({\frac{N}{d+1}}\right)^{1/2},
\end{equation} 
where the average is not over time but over all observables $A$ with a fixed spectrum but a randomly chosen eigenbasis\footnote{Here and later in this section, by random we mean chosen with respect to the unitarily invariant Haar measure.}. Hence also $\av{\av{D_A( \rho(t), \omega)}_T}_A \leq \frac{1}{2} ({{N}/({d+1})})^{1/2} \ll 1$ for any $T$. 

To make this situation more interesting, we can consider all observables for which the initial state is an eigenstate, which are typically out of equilibrium initially. For such observables, we find 
\begin{equation} 
\av{D_A( \rho(t), \omega)}_A \leq D_{\rho(0)}(\rho(t), \omega) + \frac{1}{2} \left({\frac{N}{d-1}}\right)^{1/2}
\end{equation} 
and, hence, using the result in the previous section, 
\begin{equation} 
\av{\av{D_A( \rho(t), \omega)}_T}_A \leq 
c \left(\frac{a}{\sigma_E T } + \delta \right)^{1/2}+ \frac{1}{2} \left({\frac{N}{d-1}}\right)^{1/2},
\end{equation} 
which will generally yield a very fast equilibration time.

Similar fast equilibration is obtained in Ref.\ \cite{Goldstein2015}, which considers a projector $P_{\mathrm{neq}}$ onto a subspace of non-equilibrium states of dimension $d_{\mathrm{neq}} \ll d$, and initial states within a narrow energy band with an exponentially increasing density of states characterized by inverse temperature $\beta$. When this projector is chosen at random within the energy band, any initial state leaves this space of non-equilibrium states very fast. In particular, \begin{equation}
\av{\tr(\rho(t)P_{\mathrm{neq}} )}_T \lesssim \frac{2\pi \beta}{ T}
\end{equation}
for all $T$ less than 
\begin{equation}
T_{\mathrm{max}}= 2\pi \beta \min \left\{ \left(\frac{d}{d_{\mathrm{neq}}} \right)^{\frac{1}{4}}, d^{\frac{1}{6}} \right\}.
\end{equation}

\subsubsection{Random Hamiltonians} 
Another place that randomness can be included, needless to say, is the Hamiltonian. In particular, consider that the initial state and observable as well as the spectrum of the Hamiltonian are fixed, but that the eigenbasis of the Hamiltonian is chosen at random\footnote{Note that if the initial state is entirely contained in some energy window, then one can consider only the restricted Hilbert space in that window and consider a rotation of the eigenbasis only within that subspace, in which case the results below are relative to that subspace.} \cite{Brandao2012, Reimann2016, Vinayak,Masanes2013,Cramer2012}. 
In this case, not only the equilibration time but the full time-evolution can be approximated. 
Note that this paradigmatic situation is quite different from the physically more plausible
one in which a local Hamiltonian has additional
random terms, such as in models featuring many-body localisation \cite{Schreiber2015}.
In Ref.\ \cite{Brandao2012} the equilibrium of a small system interacting with a large bath is considered in this context, and it is shown that 
\begin{equation} \label{eq:brandao} 
\av{\tr(\rho_S(t) - \omega_S)^2}_H \leq \frac{|\chi|^2}{d_S d^2} + \left( \frac{|\zeta|^2}{d^2} - \frac{\gamma}{d^2}\right)^2 + O\left(\frac{1}{d_B}\right), 
\end{equation} 
where $\av{\cdot}_H$ denotes the average over Hamiltonians, $d_S$ and $d_B$ are the dimension of the system and bath respectively (with $d=d_S d_B$),  
\begin{equation}
\chi =\sum_{k=1}^{d_E} d_k e^{2 i E_k t}, \quad \zeta= \sum_{k=1}^{d_E} d_k e^{i E_k t}, \quad \gamma =\sum_{k=1}^{d_E} d_k^2,
\end{equation}
where $d_k$ is the degeneracy of the $k^{\mathrm{th}}$ energy level. {The bound given in \eqref{eq:brandao} can be straightforwardly extended to a bound on the trace-distance, which describes how well $\rho_S(t)$ can be distinguished from $\omega_S$ using any measurement\footnote{In particular $D( \rho_S(t), \omega_S) = \frac{1}{2} \|\rho_S(t) - \omega_S\|_1$  can  be understood operationally in terms of the maximal success probability $p$ of guessing correctly whether the state is $\rho_S(t)$ or $\omega_S$ (given each with equal probability) using any measurement on the system, via $D( \rho_S(t), \omega_S) = 2p-1$.}, via $\av{\|\rho_S(t) - \omega_S\|_1}_H \leq ({d_S \av{\tr(\rho_S(t) - \omega_S)^2}_H })^{1/2}$}. Related results for subsystem equilibration in the presence of a random Hamiltonian are given in Refs.\ \cite{Vinayak,Masanes2013,Cramer2012}.

The equilibration of a particular observable with respect to a random Hamiltonian is shown in Ref.~\cite{Reimann2016} to be approximately given by 
\begin{equation} 
\tr(\rho(t) A) \simeq \tr(\rho_{\rm av} A) + F(t) \big(  \tr(\rho(0) A) - \tr(\rho_{\rm av}  A) \big)
\end{equation}  
for the vast majority of times and choices of Hamiltonian, where $\rho_{\rm av}$ is the equilibrium state $\omega$ averaged over  different choices for the Hamiltonian (resulting in $\rho_{\rm av}$ being close to the maximally mixed state), and 
\begin{equation} 
F(t) = \frac{d}{d-1} \left( \left|\frac{1}{d} \sum_{j=1}^d e^{i E_j t} \right|^2 -\frac{1}{d} \right) .
\end{equation} 
Note that $F(0) =1$, and that $F(t)$ tends towards zero as $t$ increases, becoming $O(1/d)$ in the large $t$ limit in which the phases randomise.  In the physically relevant case in which the initial state lies within a microcanonical energy window of width $\Delta E$ with exponentially increasing density of states, corresponding to a thermal bath with inverse temperature $\beta$ (where $\beta \Delta E \gg1$), then 
\begin{equation} 
 F(t) \simeq \frac{1}{1+ (t/\beta)^2}, 
\end{equation} 
giving an equilibration time comparable with $\beta$, which is  very fast. Furthermore, in this case the equilibrium state is typically very close to the microcanonical state (i.e., the state thermalises as well as equilibrates). Excitingly, this  equilibration behaviour has been observed in experiments \cite{Reimann2016}. What is more, similar equilibration behaviour can be shown even when the energy eigenstates are randomly permuted rather than chosen at random \cite{Balz2017}, although in this case the system will not generally thermalise. 

\subsubsection{Random states} The final place to introduce randomness  is the initial state, for a fixed observable and Hamiltonian. As in the case of a random observable, most initial states are already equilibrated. However, interesting results can be obtained by dividing the quantum system up into a particular small subsystem of interest, and a large bath which is in a randomly chosen, or highly mixed, initial state \cite{Garcia-Pintos2015a}. The key technical result is 
\begin{equation} 
\av{(\tr(\rho(t) A) -  \tr(\omega A))^2 }_T \leq 4\pi \|A\|^2 d\, \tr(\rho(0))^2 \xi_{\frac{1}{T}}, 
\end{equation} 
where $\xi_{\frac{1}{T}} $ is the analogous function to $\eta_{\frac{1}{T}}$, but applied to the probability distribution
$p_\alpha ={|v_\alpha|}/{Q}$ where $Q=\sum_\alpha|v_\alpha|$. 
We can  bound $\xi_{\frac{1}{T}} $ analogously to \eqref{eq:etabound} by
\begin{equation} 
\xi_{\frac{1}{T}} \leq  \left(\frac{a}{\sigma_G T } + \delta \right),
\end{equation} 
introducing parameters $a,\delta>0$ which characterise the probability distribution, and denoting by $\sigma_G$ the standard deviation of  gaps with respect to this distribution. 
If $p_\alpha$ is approximately unimodal with a dense spectrum, we would expect $a \sim 1$ and $\delta \ll 1$ as before. If the state is highly mixed initially then this can give fast equilibration times. For example if  $\rho(0) = \rho_S \otimes {I_B}/{d_B}$ for a small system of dimension $d_S$ (on which the observable $A$ acts) interacting with a  maximally mixed bath of dimension $d_B$, then 
\begin{equation} \label{eq:mixedbath}
\av{(\tr(\rho(t) A) -  \tr(\omega A))^2 }_T \leq 4\pi \|A\|^2 d_S \left(\frac{a}{\sigma_G T } + \delta \right),
\end{equation} 
which leads to equilibration times comparable to ${a d_S}/{\sigma_G}$ and independent of the bath size.

This result can be extended to bound the equilibration time for a system interacting with a bath in the microcanonical state of width $\Delta E$  with an exponentially increasing density of states with inverse temperature $\beta$ (where $\beta \Delta E \gg1$) to get \cite{Garcia-Pintos2015a}\footnote{Note that this is a slightly simplified form of the bound given in Ref.\ \cite{Garcia-Pintos2015a} neglecting minor corrections. The version given in Ref.\ \cite{Garcia-Pintos2015a} also includes the bound  $\sigma_G^2 \geq {({Q \|A\|})^{-1} | \tr([\rho(0), H],H] A) |}$ which can be substituted in place of $\sigma_G$ and is simpler to compute.} 
\begin{eqnarray} 
\av{(\tr(\rho(t) A) -  \tr(\omega A))^2 }_T \lesssim & 4\|A\|^2 \left( \pi d_S e^{\beta \|H_S\| + (1+\sqrt{d_S}) K \beta \|H_I\|}\left(\frac{a}{\sigma_G T } + \delta \right) + \frac{18}{K^2} \right),
\end{eqnarray} 
where $H_S$ and $H_I$ are the system and interaction Hamiltonians respectively, and $K$ is an arbitrary constant. Note that the bath Hamiltonian and bath dimension do not  feature. Although this result and (\ref{eq:mixedbath}) refer to mixed initial states of the bath, very similar results can be obtained for randomly chosen pure initial states of the bath.

\section{Summary and Outlook}

We have discussed the problem of equilibration times in closed quantum systems from a heuristic as well as a rigorous point of view. From the point of view of many-body physics, it seems highly plausible that generic, complex many-body systems  equilibrate in a time that depends only weakly on the system-size. However, while quite a few results are available in the case of integrable systems \cite{Calabrese2006,Calabrese2007,Eisler2007,Rigol2007,Cramer2008,Cramer2008a,Barthel2008,Flesch2008,Calabrese2011,Caux2013,Gaussification}, whose equilibration behaviour follows a power-law, very little can be said generally about this equilibration time in strongly interacting systems; in particular, it is not yet clear in which way which concrete physical properties influence how quickly a system equilibrates.  Numerical studies on classical computers are out-of-reach for these questions since large systems have to be simulated for long times. Therefore this question is an ideal use-case for forthcoming quantum simulators in which Hamiltonians and initial states can be controlled reliably.

While the heuristic discussion provided significant evidence that locally interacting many-body systems indeed
equilibrate, the given arguments are not mathematically rigorous. 
The rigorous results presented in the subsequent section, on the other hand, generally provide fairly weak general bounds on equilibration times, or extremely fast equilibration when some part of the setup is chosen at random. 
It is thus highly desirable to bridge the two worlds by incorporating general properties of many-body systems as assumptions to obtain stronger, yet rigorous bound on equilibration times which do not rely on randomness. 
We hope that this book chapter can provide a starting point and as an invitation for further researchers to study this interesting 
and important problem.

\bigskip


\paragraph*{Acknowledgements}
We would like to thank Lea F. Santos, C. Gogolin, and P. Reimann for comments on an earlier draft.
H.~W. and J.~E. acknowledge funding from the Studienstiftung des Deutschen Volkes, the ERC (TAQ),
the DFG (EI 519/14-1, EI 519/7-1, CRC 183), and the Templeton Foundation.
T.~R.~O. is supported by the Brazilian National Institute for Science and Technology of Quantum Information (INCT-IQ) and the National Counsel of Technological and Scientific Development (CNPq).

 \vspace*{-.2cm}
\begin{spacing}{0.15}
\setlength{\bibsep}{0.25pt}
\bibliographystyle{apsrev4-1}
\small{
%

}
\end{spacing}

%
%










\end{document}